# Assessment of absorbed power density and temperature rise for nonplanar body model under electromagnetic exposure above 6 GHz

**Yinliang Diao[1,2], Essam A. Rashed[1,3], and Akimasa Hirata[1,4]**

[1]Department of Electrical and Mechanical Engineering, Nagoya Institute of Technology, Nagoya 466-8555, Japan

[2]College of Electronic Engineering, South China Agricultural University, Guangzhou 510642, China

[3]Department of Mathematics, Faculty of Science, Suez Canal University, Ismailia 41522, Egypt

[4]Center of Biomedical Physics and Information Technology, Nagoya Institute of Technology, Nagoya 466-8555, Japan

## Abstract

The averaged absorbed power density (APD) and temperature rise in body models with nonplanar surfaces were computed for electromagnetic exposure above 6 GHz. Different calculation schemes for the averaged APD were investigated. Additionally, a novel compensation method for correcting the heat convection rate on the air/skin interface in voxel human models was proposed and validated. The compensation method can be easily incorporated into bioheat calculations and does not require information regarding the normal direction of the boundary voxels, in contrast to a previously proposed method. The APD and temperature rise were evaluated using models of a two-dimensional cylinder and a three-dimensional partial forearm. The heating factor, which was defined as the ratio of the temperature rise to the APD, was calculated using different APD averaging schemes. Our computational results revealed different frequency and curvature dependences. For body models with curvature radii of >30 mm and at frequencies of >20 GHz, the differences in the heating factors among the APD schemes were small.

## 1. Introduction

With the advancement of technology, there has been a trend of adopting higher frequencies to achieve larger bandwidths and hence higher data transfer rates. Additionally, fifth-generation mobile communications allow the use of frequency bands above those employed for current wireless communications (typically <6 GHz) (Lee *et al* 2018, Imai *et al* 2015). Exposure to electromagnetic (EM) radiation in this frequency range may potentially cause adverse heating effects on biological tissues. Limits for exposure to radio frequency (RF) EM fields in the range of 100 kHz to 300 GHz have been established by the International Commission on Non-Ionizing Radiation Protection (ICNIRP) (ICNIRP 2020) and the IEEE International Committee on Electromagnetic Safety (IEEE ICES) Technical Committee 95 (Bailey *et al* 2019). Both the guidelines and standards specify not-to-be-exceeded internal EM doses, which were determined according to the threshold for established health effects with reduction (safety) factors. ICNIRP calls this limit the basic restriction (BR), and IEEE refers to it as the dosimetric reference limit (DRL).

Above 6 GHz, the heating effect occurs predominantly in the superficial tissue, particularly in the skin (Ziskin *et al* 2018, Hirata *et al* 2019). With an increase in the frequency, the EM power absorption in the body and the resultant temperature rise become more superficial. For RF exposure above 6 GHz, the BR is specified in terms of the absorbed power density (APD) in the ICNIRP-2020 guideline, and the incident power density (IPD) corresponding to the APD is used as the reference level (ICNIRP 2020). The equivalent DRL is defined as the epithelial power density in IEEE C95.1 (Bailey *et al* 2019). For convenience, the term "APD" is used for both the APD and the epithelial power density herein.

Morimoto *et al* (2016) evaluated the specific absorption rate (SAR) and temperature rise in human head models up to 30 GHz. Strong correlations between the peak averaged SAR and temperature rise were observed at frequencies up to 3–4 GHz, and weaker correlations were observed at higher frequencies. Dosimetric studies for frequencies above 6 GHz have been conducted for both near- and far-field exposure scenarios (Alekseev *et al* 2008, Kanezaki *et al* 2010, Wu *et al* 2015, Sasaki *et al* 2017, Li *et al* 2019, He *et al* 2018, Xu *et al* 2017, Colombi *et al* 2018). Theoretical solutions to EM and bioheat problems for a two-dimensional (2D) layered body model were derived, and Monte Caro analyses were performed for different tissue thicknesses (Sasaki *et al* 2017, Li *et al* 2019). Samaras and Kuster (2019) computed the transmittance coefficients for a 2D exposure scenario using the exact solution for a simple layered body model. Nakae *et al* (2020) revealed that the normal component of the IPD correlates well with the skin-temperature rise, regardless of the incident angle. Several studies have been performed on the correlation between the power-density averaging area and the skin surface temperature (Hashimoto *et al* 2017, Neufeld *et al* 2018, Funahashi *et al* 2018). Funahashi *et al* (2018) reported that the averaged APD over 4 cm$^2$ exhibited a strong correlation with the skin-temperature rise in the frequency range of 30–300 GHz and was reasonable and conservative at frequencies as low as 10 GHz. Kageyama *et al* (2019) developed an exposure system and measured the temperature rise in forearm skin exposed to a focused beam generated by a lens antenna at 28 GHz. In the foregoing studies, cubic models with planar surfaces were generally adopted for the assessment of millimeter-wave exposure, in contrast to dosimetric studies for frequencies below ~10 GHz (Hirata *et al* 2002, Conil *et al* 2011, Dimbylow *et al* 2008, Wu *et al* 2011), where anatomically accurate body models have been widely used. The effects of the curvature of the body surface and the internal tissue composition on the heating factors remain unclear. These effects are difficult to evaluate directly via theoretical solutions, and numerical approaches are preferred. Additionally, current product safety standards have not yet provided detailed calculation schemes for the averaged APD, particularly for human models with curved surfaces used in numerical dosimetry. Open questions include the following: i) Should the averaging area be parallel to the grid axes in the finite-difference time domain (FDTD) method or bent along the curved body surface? ii) What are the integration limits for the curved body surface? For details regarding additional studies, the reader is referred to previous reviews (Foster *et al* 2016, Hirata *et al* 2019).

Considering the rationale for setting the limit, the APD should be linked to the maximum temperature rise. There are several major uncertainty sources in the thermal analyses, particularly for models with curved surfaces. The finite-difference (FD) approach has been commonly used in thermal analysis with voxelized human models. Laakso (2009) investigated the effect of the spatial resolution on the SAR and temperature in head models under RF plane-wave exposure. In general, the error in SAR computation results in that of the temperature rise. The error in the temperature computation is generally smaller than that in the SAR computation, because of the heat diffusion. One disadvantages of the FD analysis is that the model surface is discretized into small cubes, resulting in a large surface area. This increases the heat transfer from the model, resulting in underestimation of the temperature rise, if the measured heat-transfer coefficient is directly used as a boundary condition. A few computational schemes have been proposed for improving the accuracy of the heat flux at the model surface (Laakso 2009, Samaras *et al* 2006, Neufeld *et al* 2007). Neufeld *et al* (2007) proposed a conformal scheme for correcting the heat flux. However, this method requires accurate knowledge of the normal directions for each boundary voxel. A simple correction method was adopted by Laakso (2009); the heat convection rate for the skin voxel $H$ was corrected as $H/\sqrt{n}$, where $n$ represents the number of neighboring air voxels. However, the total surface area of a voxel model was still slightly overestimated (Laakso and Hirata 2011).

The objective of this study was to reliably assess the averaged APD and temperature rise for body models with nonplanar surfaces at frequencies above 6 GHz. Different calculation schemes for the averaged APD were proposed and evaluated. We proposed a local compensation method for the heat convection at the model boundary in bioheat calculations. The heating factors (ratios of the surface temperature rise to the APD) for the nonplanar body model were evaluated to investigate the discrepancies caused by the different curvature radii and frequencies.

## 2. Models and numerical methods

### 2.1 Models

Two models were adopted for evaluation of the APD and temperature rise. The first was a 2D cylindrical multilayer model, as shown in Fig. 1 a). This model comprised three types of tissues: skin (thickness of 1.4 mm), fat (4 mm), and muscle. Different outer radii of the cylindrical models (i.e., 20, 30, 40, and 50 mm) were considered. The other model was a three-dimensional (3D) partial forearm model extracted from the XCAT phantom (Segars *et al* 2010), as shown in Fig. 1 b). The forearm model consisted of seven types of tissues (skin, fat, muscle, blood, tendon, and cortical and cancellous bones). For all the models, a spatial resolution of 0.1 mm was adopted.

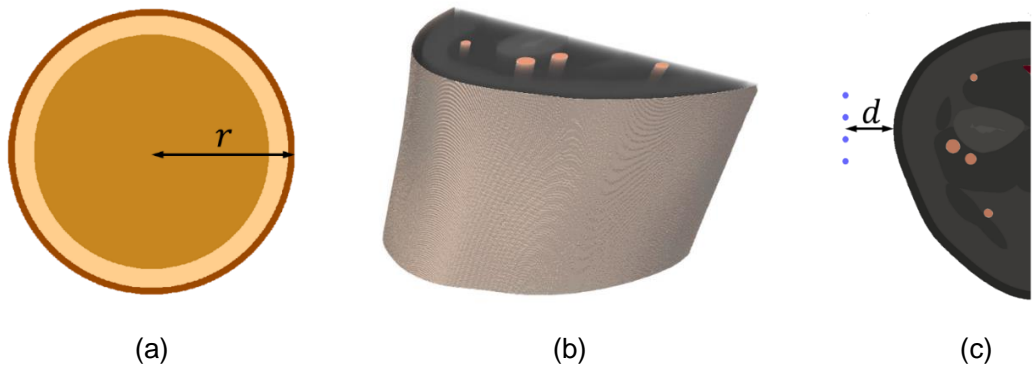

(a)          (b)          (c)

Fig. 1 Models used for numerical assessments. (a) 2D cylindrical multilayer model; (b) 3D partial forearm model with a height of 40 mm; (c) position of the source at a distance *d* from the forearm model.

### 2.2 EM calculations

For EM calculations, the FDTD method (Taflove and Hagness 2005) was used. A 15-layer convolutional perfectly matched layer (Roden and Gedney 2000) was adopted to truncate the simulation domain. For 2D analysis, both transverse-electric (TE) and transverse-magnetic (TM) plane incident waves were considered as radiation sources. For 3D analysis, a 4 × 1 half-wave dipole antenna array was used. The antenna array was located in the front of the forearm, as shown in Fig. 1 c). Different distances ($d$) between the antenna and the forearm, i.e., 5, 10, 15, 20, 30, and 40 mm, were considered. The four dipole elements were fed by delta-gap voltage sources with the same amplitudes and phases. The local SAR was calculated as the average of the electric field components, which were defined on the edges of the target voxel.

The simulated frequencies ranged from 6 to 60 GHz for the 2D analyses. The dielectric properties were obtained from the work of Gabriel, Lau, and Gabriel (1996) and are presented in Table I. For the 3D analysis, the working frequency was set as 28 GHz. The forearm model comprised seven tissues, and their dielectric properties at 28 GHz are presented in Table II.

TABLE I. Dielectric Properties of the Tissues of the Multilayer Model

| Tissue | Conductivity $\sigma$ [S/m] | | | | | | Relative permittivity $\epsilon_r$ | | | | | |
|---|---|---|---|---|---|---|---|---|---|---|---|---|
| Freq [GHz] | 6 | 10 | 20 | 30 | 45 | 60 | 6 | 10 | 20 | 30 | 45 | 60 |
| Skin | 3.89 | 8.01 | 19.2 | 27.1 | 33.4 | 36.4 | 34.9 | 31.3 | 22.0 | 15.5 | 10.4 | 7.98 |
| Fat | 0.31 | 0.59 | 1.26 | 1.79 | 2.39 | 2.82 | 4.94 | 4.60 | 4.00 | 3.64 | 3.32 | 3.13 |
| Muscle | 5.20 | 10.6 | 24.7 | 35.5 | 46.1 | 52.8 | 48.2 | 42.8 | 31.0 | 23.2 | 16.5 | 12.9 |

TABLE II. Dielectric Properties of the Tissues of the Forearm Model at 28 GHz

| Tissue | Conductivity $\sigma$ [S/m] | Relative permittivity $\epsilon_r$ |
|---|---|---|
| Skin | 25.8 | 16.6 |
| Fat | 1.70 | 3.70 |
| Muscle | 33.6 | 24.4 |
| Bone (Cortical) | 4.94 | 5.17 |
| Bone (Cancellous) | 8.87 | 7.51 |
| Blood | 37.0 | 23.9 |
| Tendon | 23.6 | 13.9 |

## 2.3 Calculation schemes for averaged APD

There are two definitions of the spatial-average APD in ICNIRP-2020 (ICNIRP 2020). The first is as follows:

$$S_{\text{ab}} = \iint_A dx dy \int_0^{z_{\max}} \rho(x,y,z) \cdot \text{SAR}(x,y,z) dz / A, \quad (1)$$

where $z = 0$ corresponds to the body surface, $z_{\max}$ encloses most of the power deposition in the body, and $A = 4$ cm$^2$ represents the averaging area. The other definition is as follows:

$$S_{\text{ab}} = \iint_A \text{Re}[\boldsymbol{S}] \cdot d\boldsymbol{s}/A = \iint_A \text{Re}[\boldsymbol{E} \times \boldsymbol{H}^*] \cdot d\boldsymbol{s}/A, \quad (2)$$

where $d\boldsymbol{s}$ represents the integral variable vector normal to the body surface. In IEEE C95.1 (Bailey et al 2019), the APD is defined as the EM power flow through the epithelium per unit area under the stratum corneum. This definition is equivalent to (2). In both the guidelines/standards, the APD should be averaged over a square area of 4 cm$^2$ at frequencies between 6 and 300 GHz. This averaging area generally provides good estimations of the surface temperature rise (Funahashi *et al* 2018, Hashimoto *et al* 2017).

For practical implementation of the averaged APD in voxel models, the definition of (1) was adopted. Moreover, we developed four different calculation schemes for the spatial-average APD for nonplanar models, as illustrated in Fig. 2. The bounds of the integration volumes are represented by red polygons. In Figs. 2 a) and b), the upper bound L1 is parallel to the grid axis. In Figs. 2 c) and d), L1 is bent along the skin surface. In Figs. 2 a) and c), L2 and L3 are parallel to the grid axis, whereas in Figs. 2 b) and d), L2 and L3 are parallel to the internal electric field gradients. The lower bound of the integration volume, i.e., L4, is defined as the contour where the electric field strength is 1/1000 of the maximum value in the integration volume.

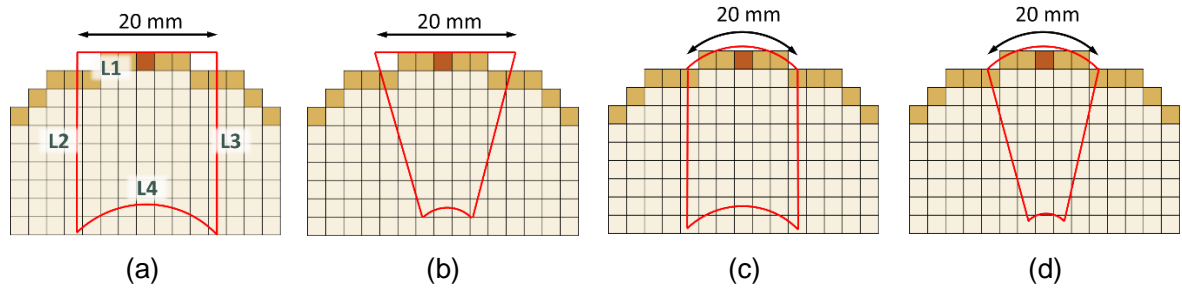

(a)　　　　　　　(b)　　　　　　　(c)　　　　　　　(d)

Fig. 2. Definitions of the integration volumes for the different APD calculation schemes. (a) Scheme 1; (b) scheme 2; (c) scheme 3 and (d) scheme 4.

### 2.4 Bioheat calculation

The widely used bioheat equation (Pennes 1948) was employed for calculation of the temperature rise inside the human body model. Under the assumption of the steady-state condition, the following equation was used for the temperature rise $\Delta T$:

$$\nabla \cdot (k \nabla \Delta T) - B \Delta T + Q_v = 0, \tag{3}$$

where $k$ [W/(m·°C)] represents the thermal conductivity; $B$ [W/(m³·°C)] is a coefficient related to the blood perfusion rate; and $Q_v$ [W/m³] represents the power-loss density, which is related to the SAR by $Q_v = \text{SAR} \cdot \rho$, where $\rho$ [kg/m³] represents the tissue mass density. The temperature rise $\Delta T$ is defined at the center of each voxel.

The Neumann boundary condition in (4) was applied to the boundary of the model:

$$k \frac{\partial \Delta T}{\partial n} + H \Delta T = 0, \tag{4}$$

where $H$ [W/(m²·°C)] represents the heat convection rate from the skin to the surrounding air. In this study, $H$ was set as 8 W/(m²·°C) (Hirata *et al* 2007). The thermal parameters used in this study were identical to those employed by Hirata, Fujiwara, and Shiozawa (2006).

## 2.5 Compensation method for heat convection rate at air/skin interface

For the implementation of the Neumann boundary condition, the heat flux through the boundary of the voxel via convection must equal the heat reaching that voxel from its neighboring voxels via conduction (Bernardi *et al* 1998). The total heat flux through a boundary voxel is given as $H \cdot S$, where $S$ represents the effective area of the air/skin interface. For a voxel belonging to a planar surface parallel to the voxel surfaces, $S = \Delta^2$, whereas in stepped boundaries, $S > \Delta^2$. To reduce the uncertainties in bioheat calculations, an accurate estimation of the surface area is required. Mullikin and Verbeek (1993) proposed a simple method for estimating surface area of 3D binary objects. This method assigns surface-area weights to each boundary voxel. The boundary voxels can be classified into five types: $S_n, n = 1, 2, \ldots, 5$, where $n$ represents the number of voxel faces exposed to the background. Fig. 3 a) shows a local boundary region containing different types of boundary voxels. The total surface area of the model can be estimated as follows:

$$A = \left( \sum_{n=1}^{5} w_n \cdot N_n \right) \cdot \Delta^2, \tag{5}$$

where $N_n$ represents the number of voxels of type $S_n$, and $w_n$ represents the weight for voxel type $s_n$. The following optimized $w_n$ values were reported: $w_1 = 0.8940$, $w_2 = 1.3409$, and $w_3 = 1.5879$ (Mullikin and Verbeek 1993). If the spatial resolution is sufficiently high, there is no obvious deviation of the boundary region from a plane. In such regions, only three types of voxels ($S_{1-3}$) exist. Voxels of type $S_{4-5}$ can occur on sharply curved surfaces. For type $S_{4-5}$, the weights $w_4 = 2$, and $w_5 = 8/3$ were presented by Mullikin and Verbeek (1993). Nonetheless, for a sphere model, there are significantly less $S_{4-5}$ voxels than $S_{1-3}$ voxels; therefore, the weights of the $S_{4-5}$ voxels are insignificant. A simple and straightforward compensation method involves adopting a factor $w_n$ for voxel type $S_n$. Consequently, the heat flux through the boundary voxel becomes $H w_n \Delta^2$. Thus, the total heat flux through the body surface can be accurately estimated.

However, particular attention should be paid to local stepped regions, as shown in Figs. 3 b)–d). In these locally planar regions, only single type of boundary voxel exists. With the foregoing compensation method, the local effective surface area would be underestimated as $w_n < \sqrt{n}$. To resolve this issue, we propose a local correction method using a 3 × 3 × 3 moving cube, which is centered at the target boundary voxel (outlined in red in Fig. 3). Assuming that $N'_n$ represents the total number of voxels of type $S_n$ within the moving cube, if there is only one type of boundary voxel within the cube (i.e., $N'_n = N'_A$, $N'_A$ is the total number of boundary voxels), the heat transfer through the central voxel is compensated as $H\sqrt{n}\Delta^2$. Thus, the heat convection rate in stepped regions, as shown in Figs. 3 b)–d), is corrected.

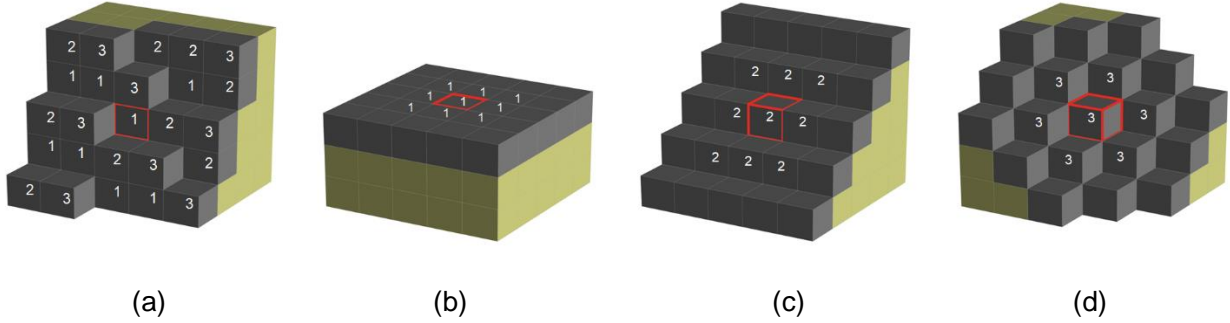

Fig. 3. Air/skin interface for voxel models. Black voxels indicate boundary voxels; yellow voxels indicate internal tissues. The boundary voxels are labeled according to the voxel type; the center voxels are outlined in red. The region in (a) contains different types of boundary voxels. The regions in (b) to (d) contain one type of boundary voxel. The effective surface areas for the central voxels are (b) $\Delta^2$, (c) $\sqrt{2}\Delta^2$, and (d) $\sqrt{3}\Delta^2$.

## 3. Validation of compensation method for bioheat calculation

To validate the proposed compensation method for the heat flux at the air/skin interface, we considered a homogeneous sphere model with radius of 60 mm. The heat conduction rate was set as 1.0 W/(m·°C), and a 200-W/m³ heat source was uniformly distributed inside the sphere. The heat convection rate between the skin and air was set as 8 W/(m²·°C). The analytical solution to this problem is $\Delta T(r) = -200r^2/6 + 0.62$ °C, where $r$ represents the radial distance.

The calculation results of the conformal and proposed compensation methods are presented in Table III and Fig. 4. Implementation of the conformal method requires knowledge of the normal directions for each boundary voxel, which were obtained using a 3D Sobel operator in this study. As indicated by Table III, the total surface area was 2.2% larger than the real surface area for the conformal method. This resulted in an underestimation of the temperature rise, with a mean value of 0.4886 °C for the sphere surface, compared with the exact solution of 0.5 °C. The $\Delta T$ calculated using the proposed method was almost identical to the exact solution, as shown in Table III and Fig. 5. Moreover, compared with the conformal method, the proposed method provided a smaller standard deviation of $\Delta T$ for the boundary voxels.

TABLE III. Comparison of Thermal Calculation Results Obtained using Different Compensation Methods.

| | Surface area [mm$^2$] | Max $\Delta T$ [°C] | Mean surface $\Delta T$ [°C] | SD of surface $\Delta T$ [°C] |
|---|---|---|---|---|
| Exact solution | 4.523 × 10$^4$ | 0.62 | 0.5 | 0.0 |
| Conformal method | 4.623 × 10$^4$ (+2.2%) | 0.6069 (−2.11%) | 0.4886 (−2.3%) | 0.001178 |
| Proposed method | 4.521 × 10$^4$ (−0.04%) | 0.6179 (−0.34%) | 0.4997 (−0.06%) | 0.001054 |

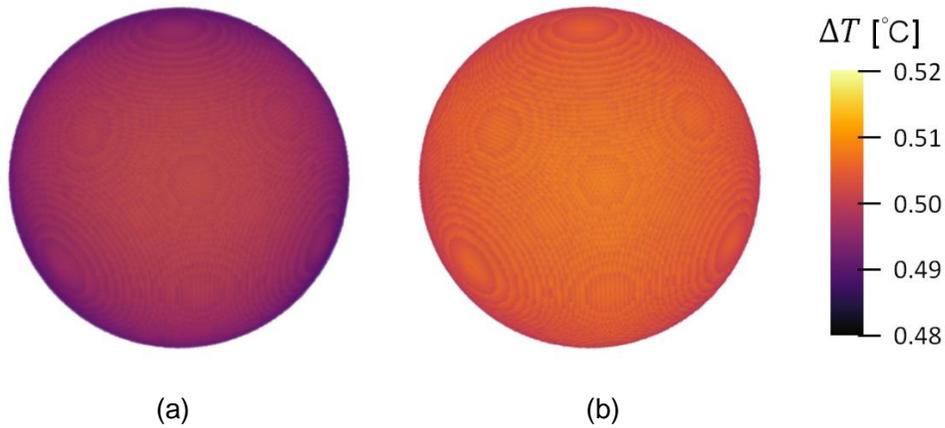

(a)      (b)

Fig. 4 Distributions of the temperature rise on the surface of a homogeneous sphere. The boundary condition was implemented using (a) conformal and (b) proposed methods.

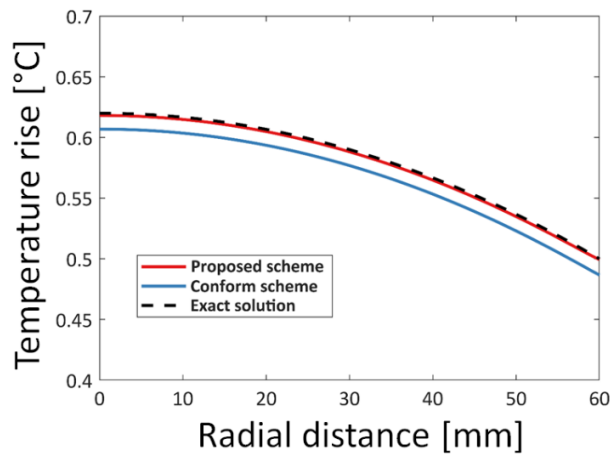

Fig. 5 Temperature-rise distributions along the sphere radius for different compensation methods.

An ellipsoidal model, as shown in Fig. 6 a), was also adopted for validation. The three principal semi-axes of the ellipsoid were set as $a = 30$, $b = 50$, and $c = 100$ mm. The thermal parameters were identical to those used for the spherical model. Spatial resolutions of 1, 0.5, and 0.25 mm were employed. The ellipsoid was tilted around the y-axis in steps of 15°. The calculated maximum temperature rises inside the ellipsoids for different spatial resolutions and tilt angles are presented in Fig. 6 b). As shown, the maximum temperature rises inside the models and on the surfaces were almost independent of the tilt angle, and the relative standard deviations caused by the tilt angle were smaller than ~0.3%. The discrepancies of the maximum temperature rises inside the models and on the model surfaces from those obtained via the MATLAB Partial Differential Equation Toolbox[TM] (version R2020a) (MathWorks 2020) using the finite-element method (FEM) were within ~0.8% and ~0.5%, respectively.

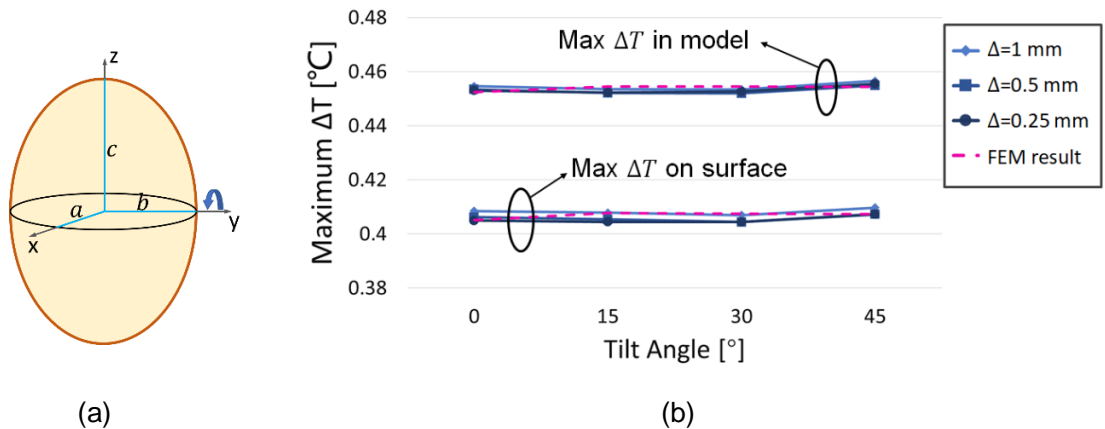

(a)　　　　　　　　　　　　　　　　　　(b)

Fig. 6 Validation of the proposed compensation method using prolate ellipsoidal models. (a) Ellipsoidal model (original position), with an arrow indicating the rotation direction; (b) maximum temperature rises inside the model and on the model surface.

## 4. APD and temperature rise for nonplanar models

### 4.1 2D multilayer model

The calculated SAR and temperature rise in the 2D multilayer models for TE and TM incident waves are presented in Figs. 7 a) and b), respectively. As shown, the EM power depositions were distributed superficially above 10 GHz. The peak local SARs for the TM cases were lower than those for the TE cases. This is because for the TE cases, the electric fields were parallel to the axis of the infinitely long cylindrical model. The maximum temperature rises were generally slightly larger for the TM cases than for the TE cases. As shown in Fig. 7 b), the EM wave penetrated into the cylindrical model from different incident angles, resulting in broader SAR and temperature distributions compared with the TE case. This phenomenon was attributed to Brewster's effect (Li *et al* 2019).

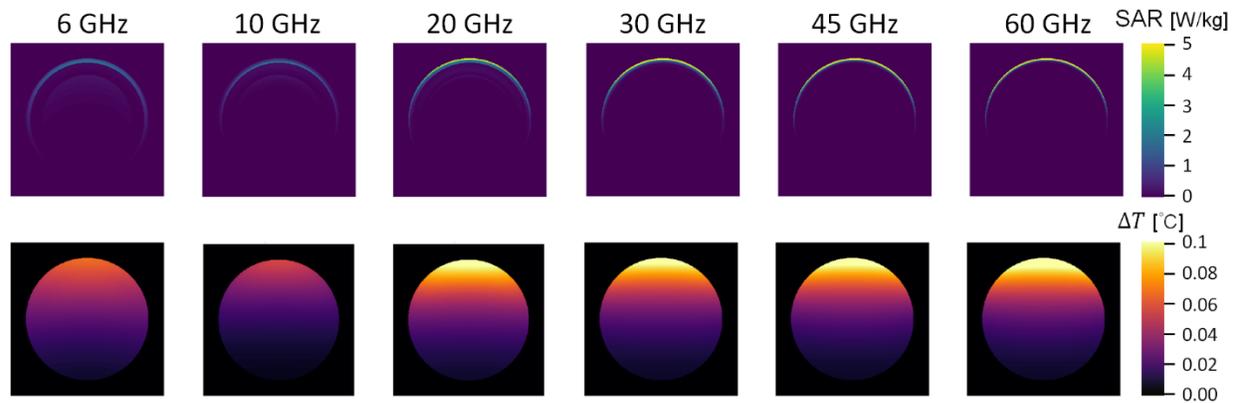

(a)

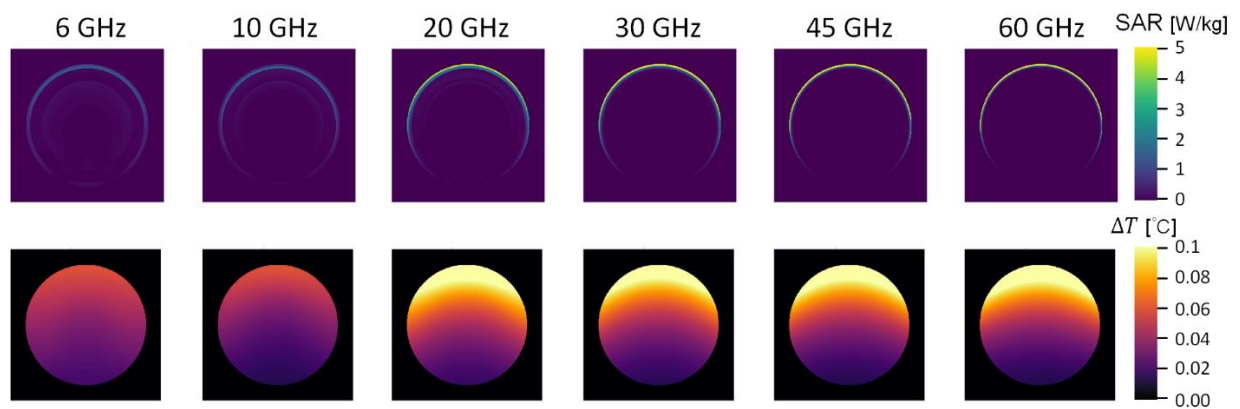

(b)

Fig. 7. Distributions of the SAR and temperature rise on the 2D multilayer models with a 20-mm radius for (a) TE and (b) TM incident waves.

In Fig. 8, the orange regions indicate the integration volumes for different APD calculation schemes at various frequencies. For the cylindrical model with a 20-mm radius, increasing the depth to enclose all the power transmitted through L1 is impossible at 6 and 10 GHz. Therefore, the lengths of L2 and L3 are set as 20 mm at 6 and 10 GHz for the 20-mm-radius model. This setting is in accordance with the definition of the 10-g spatial-average SAR, where the side length of the equivalent integration cube is approximately 20 mm (Hirata *et al* 2019).

The calculated heating factors for different APD schemes are presented in Figs. 9 and 10 for TE and TM incident waves, respectively. As shown, the heating factors for the TM cases were slightly larger than those for the TE cases. The largest differences in the heating factor between the TE and TM cases was ~20%, for scheme 4 of the 20-mm-radius model at 6 GHz. The differences decreased with the increasing frequency and curvature radius. For frequencies above ~20 GHz and curvature radii larger than ~30 mm, the differences in the heating factor between the TE and TM cases were smaller than ~6%.

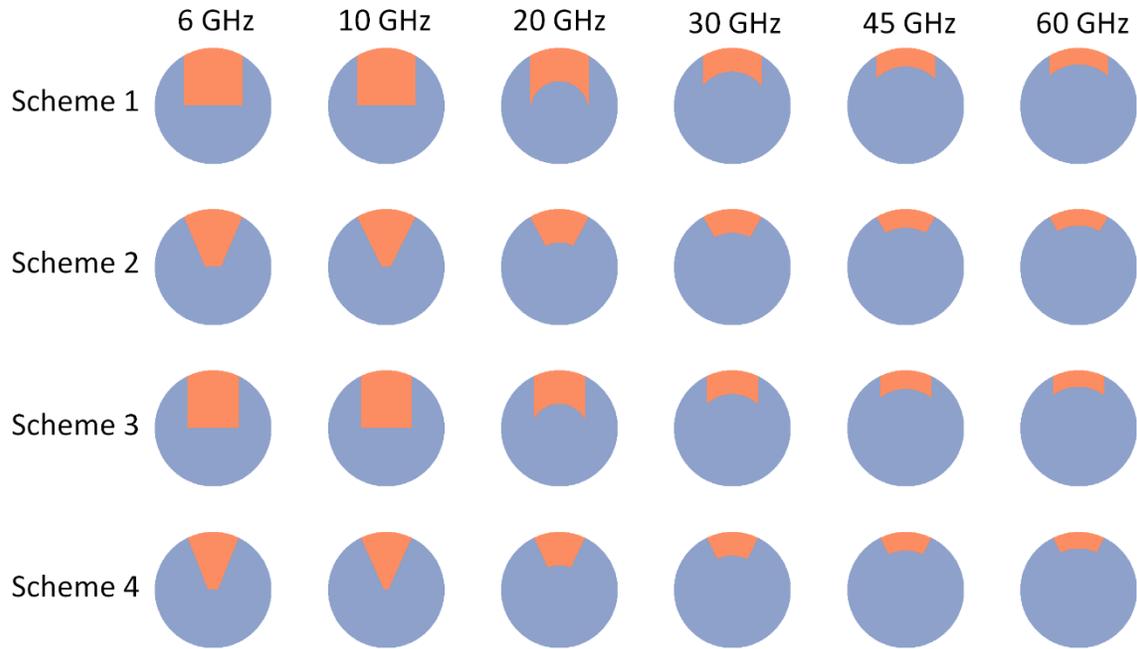

Fig. 8. Integration volumes for different APD schemes. The radius of the cylindrical models is 20 mm. Orange shapes indicate different integration volumes.

The effect of the curvature radius on the heating factor was rather small. For the TE cases, the largest difference in the heating factor between the 20- and 50-mm-radius models was ~10% (at 6 GHz). For the TM cases, the largest difference was ~20% (at 10 GHz).

Above ~20 GHz, the heating factors for the APD were almost independent of the frequency. Reductions in the heating factors below 10 GHz were observed for most schemes. For a smaller curvature radius, the EM power loss was more likely to be transferred to the air from the closest skin surface via heat convection than to contribute to the temperature rise. For convenience, we denote the heating factor evaluated via scheme $i$ as $HF_i$. At 6 GHz, $HF_1$ and $HF_3$ were ~20% smaller than those at higher frequencies; while $HF_2$ and $HF_4$ were ~13% and ~5% smaller than those at higher frequencies for TE and TM incident waves, respectively.

For schemes 1 and 2, the upper bounds of the integration volume were tangent planes to the cylinders, whereas for schemes 3 and 4, the upper bounds were bent along the surface. Therefore, the integration volumes for the latter two schemes were smaller than those for schemes 1 and 2, resulting in larger heating factors. The differences between $HF_1$ and $HF_3$ and between $HF_2$ and $HF_4$ decreased with the increasing curvature radius. This is mainly attributed to the reduction in the deviation of the model surface from a plane. $HF_3$ and $HF_4$ decreased steadily with an increase in the radii, whereas $HF_1$ and $HF_2$ were relatively stable for all the radii.

In general, among the schemes examined, $HF_2$ seems to be most stable across all radii and frequencies. For a cylindrical radius of ≥30 mm, the four schemes provided comparable heating-factor results, with relative standard deviations of <~5% for frequencies of ≥20 GHz and <~10% for frequencies of ≥6 GHz.

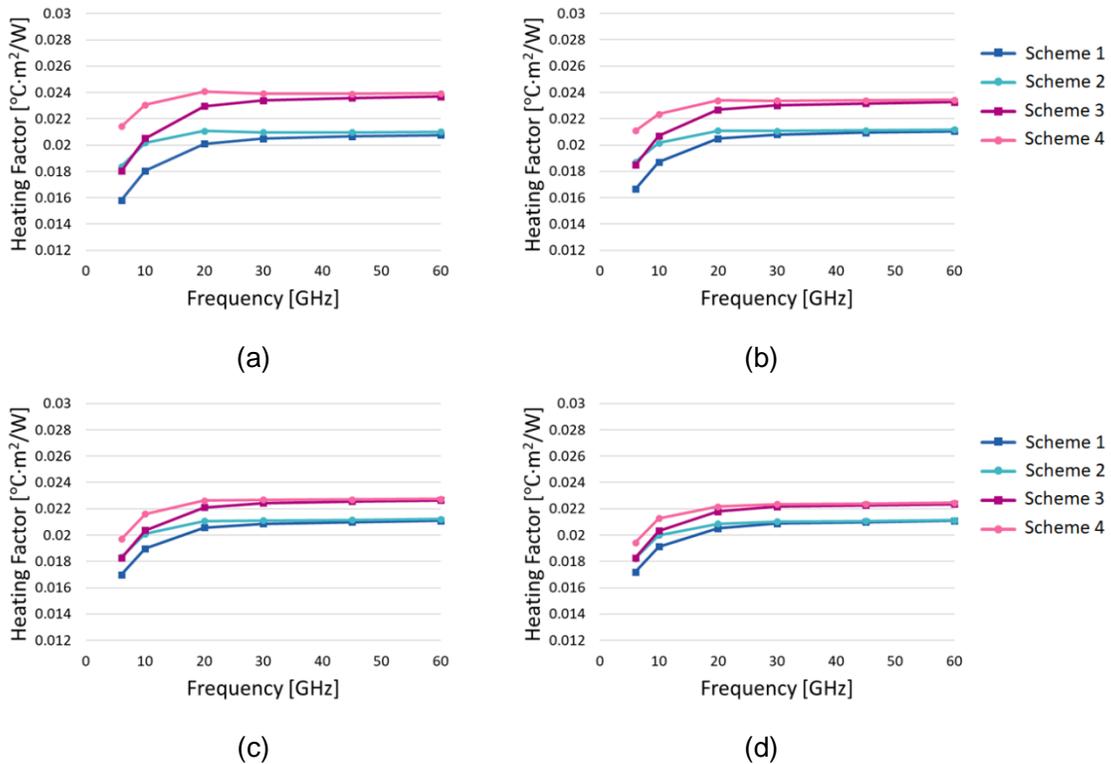

Fig. 9. Heating factors calculated using different APD schemes for 2D cylindrical models with radii of (a) 20, (b) 30, (c) 40, and (d) 50 mm, for TE incident waves.

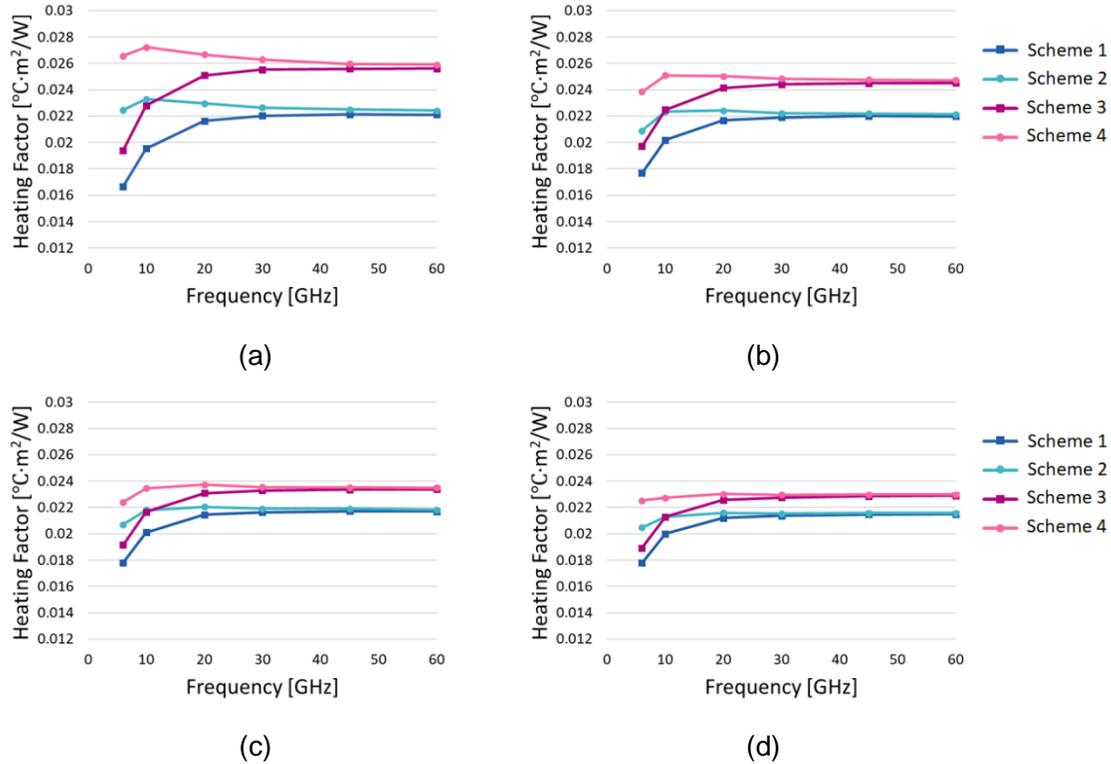

Fig. 10. Heating factors calculated using different APD schemes for 2D cylindrical models with radii of (a) 20, (b) 30, (c) 40, and (d) 50 mm, for TM incident waves.

### 4.2 3D forearm model

A partial human forearm model was adopted for 3D analysis. The temperature rises were calculated using the heat convection compensation method described in Section II. The calculated local SAR and $\Delta T$ distributions in the partial forearm model are shown in Figs. 11 a) and b), respectively.

The differences in the heating factors among the different averaging schemes were marginal and smaller than those for the 2D cylindrical models. In the 2D cases, a plane wave was used as the radiation source. Fig. 12 shows the heating factors for different distances between the antenna and the forearm. The largest heating factors were observed for $d = 5$ mm. Fluctuations in the heating factors were observed within the reactive near-field region of the antenna ($d \leqslant 15$ mm) and were mainly attributed to the oscillating peak power density in this field region (Colombi *et al* 2015). Above 15 mm, the heating factors steadily decreased with the increasing separation distance. For $d = 40$ mm, the heating factors were approximately 0.22 °C·m²/W for schemes 1 and 2 and 0.23 °C·m²/W for schemes 3 and 4. These values were slightly larger than those for 2D multilayer models with 30-mm radii (the curvature radius of the forearm was approximately 30–40 mm). This difference was attributed to the field non-uniformity and the thicker fat tissue of the forearm model. As reported by

Alekseev *et al* (2008), a thicker fat layer tends to produce a slightly larger surface-temperature rise in the skin.

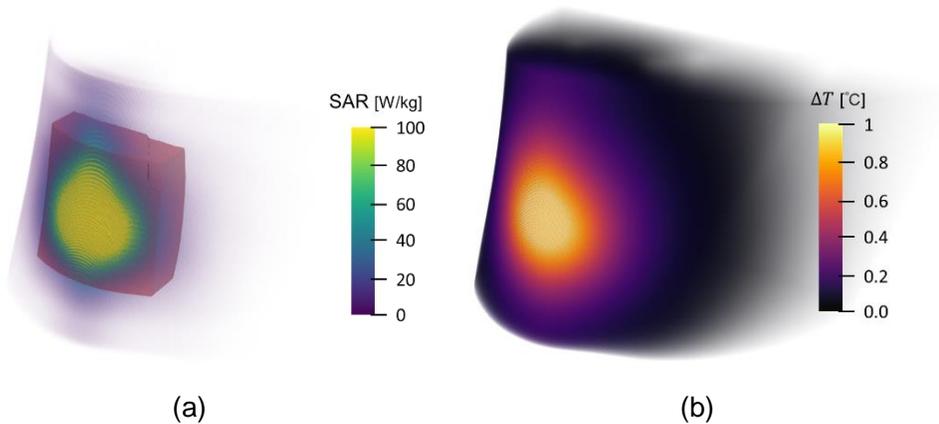

(a)          (b)

Fig. 11 Distributions of the (a) local SAR and (b) temperature rise in the partial forearm model. The antenna–forearm distance is 10 mm. The antenna accepted power is normalized to 20 dBm. The dark red box in (a) indicates the integration volume for APD averaging scheme 1.

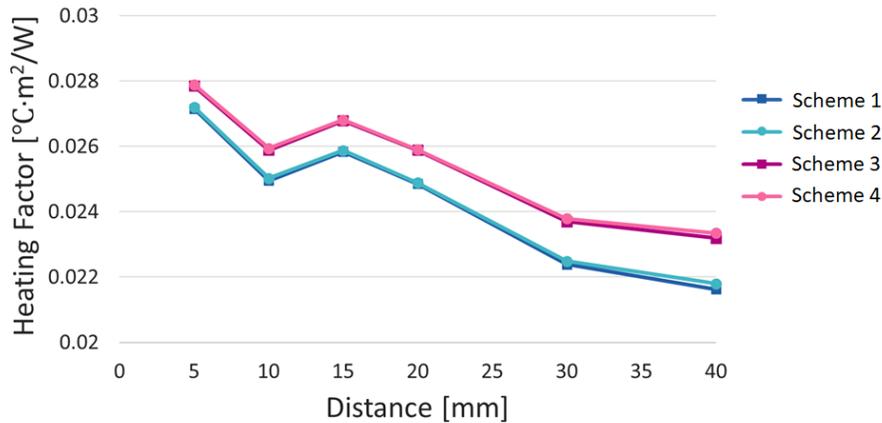

Fig. 12 Heating factors for a 4 × 1 dipole antenna array with different antenna–forearm distances.

## 5. Discussion and concluding remarks

In bioheat calculations, the boundary conditions significantly influence the calculation results. The discretization of the model surface with cubes results in a larger model surface area; hence, the temperature rise is underestimated. This issue is particularly crucial for the assessment of millimeter-wave exposure, owing to the extremely shallow penetration depth. In the study of Laakso (2009), the heat convection rate for the skin voxel, i.e., $H$, was corrected as $H/\sqrt{n}$, where $n$ represents the number of exposed faces of the voxel. However, the total surface area of the voxel model was slightly overestimated (Laakso and Hirata 2011). This problem can be solved by introducing a compensation factor $f$, which is defined as the ratio of the real body surface area (if known) to the calculated one. The compensated heat transfer from one boundary voxel then becomes $H\sqrt{n}f\Delta^2$, $f < 1$. This method is applicable to whole-body exposure scenarios, as the total heat flux through the body surface is corrected. For localized exposure scenarios, however, compensation of the whole-body surface area may still lead to underestimation of the heat flux though the local surface parallel to the grid axes (as shown in Figs. 3 b)–d)), because $f < 1$. In this paper, a new conformal compensation method was proposed for reducing the uncertainties in FD bioheat calculations for voxel models with nonplanar surfaces. This method does not require knowledge of the normal directions of the boundary voxels and can be easily incorporated into FD iterative programs with a minimal computational load. Additionally, we confirmed that the proposed method provides an accurate estimation of the temperature rise. The variation in the maximum temperature rise was observed to be smaller than ~1% for different spatial resolutions and different model rotation angles (shown in Fig. 6).

Because no clear definition of the calculation scheme for the APD is prescribed in the existing standards, four schemes were proposed, particularly for models with nonplanar surfaces. The relationships between the maximum temperature rise and the averaged APD were determined using different schemes. Compared with previous studies using planar models, comparable heating factors for the APD were observed for curvature radii larger than ~30 mm and frequencies above ~20 GHz. Above ~20 GHz, all four schemes provided stable heating factors, and the heating factors decreased below ~10 GHz. As reported by Hirata, Funahashi, and Kodera (2019), even though there may be optimized physical quantities for estimating the surface-temperature rise below 10 GHz, the APD is a good surrogate, with deviations smaller than ~20%. In general, if the curvature radius is larger than ~30 mm, the uncertainties in the heating factors for different APD schemes are small, with relative standard deviations smaller than ~5% for frequencies above ~20 GHz and within ~10% for frequencies above 6 GHz. In the 3D cases, the relative standard deviations of the heating factors are within ~3.5% for all the APD schemes.

In previous studies, slightly larger heating factors were observed for TM-like exposure using 2D infinite-slab models (Samaras and Kuster 2019, Li *et al* 2019, Nakae *et al* 2020). This phenomenon was also observed in the present study for the 2D multilayer cylindrical model, as shown in Figs. 9 and 10. However, the differences in the heating factors between the TE and TM cases were no larger than ~20%. For the 2D multilayer models, the calculated heating factors were on the order of ~0.02 °C/W·m² at frequencies of ≥10 GHz. For the 3D forearm model, when the antenna–forearm distance was shorter, the EM power was more narrowly distributed, and larger heating factors were observed. In such cases, the coupling effects between the antenna and the body are predominant under exposure to a reactive near field (Funahashi *et al* 2018). This coupling effect generally depends on the antenna design, frequency, and separation distance. If the antenna mismatch is considered, the effect is mitigated owing to the reduced output power (Colombi *et al* 2018). With an increase in the distance, the heating factors become comparable to those for incident plane waves. This is consistent with the results of Hashimoto *et al* (2017), who reported that the heating factors for beams with diameters larger than the side length of the averaging area are comparable to those for incident plane waves. In general, the calculated heating factors were in good consistence with <0.025 °C·m$^2$/W, which is suggested by ICNIRP-2020 guideline (ICNIRP 2020).

In this study, the uncertainty from the incidence angle was not considered. Owing to the rotational symmetry of the multilayer models, the results of our 2D simulation were independent of the incident angle of the exposed plane wave. Nakae *et al* (2020) also found that the heating factors for the APD are insensitive to the incident angle.

In conclusion, the averaged APD and temperature rise in body models with nonplanar surfaces under EM exposure at >6 GHz were evaluated. Different calculation schemes for the averaged APD were investigated. To reduce the uncertainties in the FD bioheat calculations caused by the staircasing artifacts at the model boundaries, we proposed a new local compensation method for correcting the heat convection rate. This compensation method was validated by analytical solution using a sphere model and by FEM solutions using prolate ellipsoidal models with different tile angles and spatial resolutions. It was demonstrated that the proposed method can significantly improve the thermal calculation at stepped boundaries, regardless of the model resolution and rotation. The APDs and temperature rises for 2D cylindrical models and a 3D forearm model were then evaluated at frequencies ranging from 6 to 60 GHz using the APD schemes and the heat convection compensation method. When the model curvature radii were larger than ~30 mm and the frequency was above ~20 GHz, the calculated heating factors agreed well with those obtained in previous studies using planar models, and the differences in the heating factors among different APD schemes were marginal.

## Acknowledgement

This work was partly supported by the Ministry of Internal Affairs and Communications, Japan.

# References

Alekseev S I, Radzievsky A A, Logani M K and Ziskin M C 2008 Millimeter wave dosimetry of human skin *Bioelectromagnetics* **29** 65–70

Bailey W H, Bodemann R, Bushberg J, Chou C, Cleveland R, Faraone A, Foster K R, Gettman K E, Graf K, Harrington T, Hirata A, Kavet R R, Keshvari J, Klauenberg B J, Legros A, Maxson D P, Osepchuk J M, Reilly J P, Tell R R A, Thansandote A, Yamazaki K, Ziskin M C and Zollman P M 2019 Synopsis of IEEE Std C95.1™-2019 "IEEE standard for safety levels with respect to human exposure to electric, magnetic, and electromagnetic fields, 0 Hz to 300 GHz" *IEEE Access* **7** 171346–56

Bernardi P, Cavagnaro M, Pisa S and Piuzzi E 1998 SAR distribution and temperature increase in an anatomical model of the human eye exposed to the field radiated by the user antenna in a wireless LAN *IEEE Trans. Microw. Theory Tech.* **46** 2074–82

Colombi D, Thors B and Törnevik C 2015 Implications of EMF exposure limits on output power levels for 5G devices above 6 GHz *IEEE Antennas Wirel. Propag. Lett.* **14** 1247–9

Colombi D, Thors B, TöRnevik C and Balzano Q 2018 RF energy absorption by biological tissues in close proximity to millimeter-wave 5G wireless equipment *IEEE Access* **6** 4974–81

Conil E, Hadjem A, Gati A, Wong M and Wiart J 2011 Influence of plane-wave incidence angle on whole body and local exposure at 2100 MHz *IEEE Trans. Electromagn. Compat.* **53** 48–52

Dimbylow P J, Hirata A and Nagaoka T 2008 Intercomparison of whole-body averaged SAR in European and Japanese voxel phantoms *Phys. Med. Biol.* **53** 5883–97

Foster K R, Ziskin M C and Balzano Q 2016 Thermal response of human skin to microwave energy: a critical review *Health Phys.* **111** 528–41

Funahashi D, Hirata A, Kodera S and Foster K R 2018 Area-averaged transmitted power density at skin surface as metric to estimate surface temperature elevation *IEEE Access* **6** 77665–74

Gabriel S, Lau R W and Gabriel C 1996 The dielectric properties of biological tissues: III. Parametric models for the dielectric spectrum of tissues *Phys. Med. Biol.* **41** 2271–93

Hashimoto Y, Hirata A, Morimoto R, Aonuma S, Laakso I, Jokela K and Foster K R 2017 On the averaging area for incident power density for human exposure limits at frequencies over 6 GHz *Phys. Med. Biol.* **62** 3124–38

He W, Xu B, Gustafsson M, Ying Z and He S 2018 RF compliance study of temperature elevation in human head model around 28 GHz for 5G user equipment application: simulation analysis *IEEE Access* **6** 830–8

Hirata A, Fujiwara O and Shiozawa T 2006 Correlation Between Peak Spatial-Average SAR and Temperature Increase Due to Antennas Attached to Human Trunk *IEEE Trans. Biomed. Eng.* **53** 1658–64

Hirata A, Funahashi D and Kodera S 2019 Setting exposure guidelines and product safety standards for radio-frequency exposure at frequencies above 6 GHz: brief review *Ann. Telecommun.* **74** 17–24

Hirata A, Watanabe H and Shiozawa T 2002 SAR and temperature increase in the human eye induced by obliquely incident plane waves *IEEE Trans. Electromagn. Compat.* **44** 592–4

Hirata A, Watanabe S, Fujiwara O, Kojima M, Sasaki K and Shiozawa T 2007 Temperature elevation in the eye of anatomically based human head models for plane-wave exposures. *Phys. Med. Biol.* **52** 6389–99

ICNIRP 2020 Guidelines for Limiting Exposure to Electromagnetic Fields (100 kHz to 300 GHz) *Health Phys.* **118** 483–524

Imai T, Kitao K, Tran N, Omaki N, Okumura Y, Sasaki M and Yamada W 2015 Development of high frequency band over 6 GHz for 5G mobile communication systems *2015 9th European Conference on Antennas and Propagation (EuCAP)* pp 1–4

Kageyama I, Masuda H, Morimatsu Y, Ishitake T, Sakakibara K, Hikage T and Hirata A 2019 Comparison of temperature elevation between in physical phantom skin and in human skin during local exposure to a 28 GHz millimeter-wave *2019 Joint International Symposium on Electromagnetic Compatibility, Sapporo and Asia-Pacific International Symposium on Electromagnetic Compatibility (EMC Sapporo/APEMC)* pp 766–9

Kanezaki A, Hirata A, Watanabe S and Shirai H 2010 Parameter variation effects on temperature elevation in a steady-state, one-dimensional thermal model for millimeter wave exposure of one- and three-layer human tissue *Phys. Med. Biol.* **55** 4647–59

Laakso I 2009 Assessment of the computational uncertainty of temperature rise and SAR in the eyes and brain under far-field exposure from 1 to 10 GHz. *Phys. Med. Biol.* **54** 3393–404

Laakso I and Hirata A 2011 Dominant factors affecting temperature rise in simulations of human thermoregulation during RF exposure *Phys. Med. Biol.* **56** 7449

Lee J, Tejedor E, Ranta-aho K, Wang H, Lee K, Semaan E, Mohyeldin E, Song J, Bergljung C and Jung S 2018 Spectrum for 5G: global status, challenges, and enabling technologies *IEEE Commun. Mag.* **56** 12–8

Li K, Sasaki K, Watanabe S and Shirai H 2019 Relationship between power density and surface temperature elevation for human skin exposure to electromagnetic waves with oblique incidence angle from 6 GHz to 1 THz *Phys. Med. Biol.* **64** 065016

MathWorks T 2020 Partial Differential Equation Toolbox (R2020a)

Morimoto R, Laakso I, Santis V De and Hirata A 2016 Relationship between peak spatial-averaged specific absorption rate and peak temperature elevation in human head in frequency range of 1-30 GHz *Phys. Med. Biol.* **61** 5406–25

Mullikin J C and Verbeek P W 1993 Surface area estimation of digitized planes *Bioimaging* **1** 6–16

Nakae T, Funahashi D, Higashiyama J, Onishi T and Hirata A 2020 Skin temperature elevation for incident power densities from dipole arrays at 28 GHz *IEEE Access* **8** 26863–71

Neufeld E, Carrasco E, Murbach M, Balzano Q, Christ A and Kuster N 2018 Theoretical and numerical assessment of maximally allowable power-density averaging area for conservative electromagnetic exposure assessment above 6 GHz *Bioelectromagnetics* **39** 617–30

Neufeld E, Chavannes N, Samaras T and Kuster N 2007 Novel conformal technique to reduce staircasing artifacts at material boundaries for FDTD modeling of the bioheat equation *Phys. Med. Biol.* **52** 4371–81

Pennes H H 1948 Analysis of Tissue and Arterial Blood Temperatures in the Resting Human Forearm *J. Appl. Physiol.* **1** 93–122

Roden J A and Gedney S D 2000 Convolutional PML (CPML): An efficient FDTD implementation of the CFS-PML for arbitrary media *Microw. Opt. Technol. Lett.* **27** 334–8

Samaras T, Christ A and Kuster N 2006 Effects of geometry discretization aspects on the numerical solution of the bioheat transfer equation with the FDTD technique *Phys. Med. Biol.* **51** N221--N229

Samaras T and Kuster N 2019 Theoretical evaluation of the power transmitted to the body as a function of angle of incidence and polarization at frequencies >6 GHz and its relevance for standardization *Bioelectromagnetics* **40** 136–9

Sasaki K, Mizuno M and Wake K 2017 Monte Carlo simulations of skin exposure to electromagnetic field from 10 GHz to 1 THz Monte Carlo simulations of skin exposure to electromagnetic field from 10 GHz to 1 THz *Phys. Med. Biol.* **62** 6993–7010

Segars W P, Sturgeon G, Mendonca S, Grimes J and Tsui B M W 2010 4D XCAT phantom for multimodality imaging research *Med. Phys.* **37** 4902–15

Taflove A and Hagness S C 2005 *Computational Electrodynamics: The Finite-Difference Time-Domain Method* 3rd edn (Boston, MA: Artech House Publishers)

Wu T, Rappaport T S and Collins C M 2015 Safe for generations to come: considerations of safety for millimeter waves in wireless communications *IEEE Microw. Mag.* **16** 65–84

Wu T, Tan L, Shao Q, Zhang C, Zhao C, Li Y, Conil E, Hadjem A, Wiart J, Lu B, Xiao L, Wang N, Xie Y and Zhang S 2011 Chinese adult anatomical models and the application in evaluation of RF exposures *Phys. Med. Biol.* **56** 2075–89

Xu B, Zhao K, Thors B, Colombi D, Lundberg O, Ying Z and He S 2017 Power density measurements at 15 GHz for RF EMF compliance assessments of 5G user equipment *IEEE Trans. Antennas Propag.* **65** 6584–95

Ziskin M C, Alekseev S I, Foster K R and Balzano Q 2018 Tissue models for RF exposure evaluation at frequencies above 6 GHz *Bioelectromagnetics* **39** 173–89